\documentstyle[prl,aps,epsfig]{revtex}
\begin{document}
\title{Probing the structure of the cold dark matter halo using ancient mica}
\author{Edward A. Baltz and Andrew J. Westphal}
\address{Department of Physics, University of California, Berkeley, CA
94720-7300}
\author{Daniel P. Snowden-Ifft}
\address{Department of Physics, Occidental College, Los Angeles, CA 90041}

\maketitle
\begin{abstract}  
Mica can store (for $>1$ Gy) etchable tracks caused by atoms recoiling from
WIMPs.  Ancient mica is a directional detector despite the complex motions it
makes with respect to the WIMP ``wind''.  We can exploit the properties of
directionality and long integration time to probe for structure in the dark
matter halo of our galaxy.  We compute a sample of possible signals in mica for
a plausible model of halo structure.
\end{abstract}
\pacs{95.35.+d,95.55.Vj,95.85.Ry,98.35.Gi}

It has been shown that 90\% of the mass of our galaxy is in a halo whose radius
is at least 50 kpc, extending far beyond the luminous matter \cite{dark}.  The
nature of this dark halo is still a subject of active research.  One promising
possibility is that this halo consists of weakly interacting massive particles
(WIMPs).  The theoretical support for this possibility is strong, both from
theories of galaxy formation and supersymmetric extensions to the standard
model \cite{susydm}.  There is a large experimental effort to detect these
particles in the halo of our galaxy.  One technique consists of searching
ancient mica for damage caused by the rare WIMP interactions \cite{mica}.  This
technique has two advantages.  The first is that the integration time can be of
order 1 Gy if suitably old mica is chosen for study.  This is unique in that
the mica samples the halo continuously along the galactic orbit.  The second is
that mica exhibits a directional sensitivity when used as a WIMP detector
\cite{etch,direction}.  In this work we describe how these two advantages can
be used in concert to probe the structure of a WIMP halo in a way that no other
dark matter experiment can.

A halo consisting of WIMPs need not have a smooth distribution in phase space
as is often assumed.  In the hierarchical clustering paradigm of halo formation
\cite{hierarchy}, small objects form earliest, and large objects are formed by
mergers between small objects.  It is far from certain that the small objects
are completely disrupted by these mergers.  We know that there are large dark
matter clumps of about 10$^8 M_\odot$ in the galactic halo, the dwarf
spheroidal galaxies.  It is plausible that there may be other such clumps
without a luminous baryonic component persisting in the halo today, providing a
substantial substructure.  An alternate scenario is that the clumps are being
tidally disrupted.  Calculations indicate that clumps can trail streamers of
tidal debris along a large fraction of their galactic orbit \cite{streamers}.

We consider a model of the halo where a fraction $f$ of the mass is in dense,
gravitationally bound clumps or streamers and the remaining fraction is
completely thermalized.  The individual clumps are modeled as isothermal
spheres with virial velocities of 40 km sec$^{-1}$ and cutoff velocities of 80
km sec$^{-1}$, though the results of this work are insensitive to the exact
values of these parameters.

A streamer looks quite similar to a clump for an observer passing through it.
We model a streamer of mass $M$ which is stretched relative to its diameter by
a factor of $\eta$ as being equivalent to a number $\eta$ of separate clumps
each with mass $M/\eta$.  If we consider a halo consisting of roughly $10^4$
clumps, each with mass $10^8 M_\odot$ like the dwarf spheroidals, and we assume
that $\eta\sim 100$ as is shown in simulations \cite{andi}, we find a halo
which is effectively made from about $10^6$ clumps of mass $10^6 M_\odot$.

We adopt the standard parameters for the isothermal component of the halo.  The
thermal velocity is 261 km sec$^{-1}$ and the cutoff velocity is 640 km
sec$^{-1}$.  The local density of the isothermal dark matter sphere is taken to
be 0.3 $(1-f)$ GeV cm$^{-3}$ \cite{susydm}.

We study two density profiles for the clump fraction of the halo, those of
Hernquist \cite{hernquist} and Navarro, Frenk, and White (NFW) \cite{navarro}.
Both are isotropic and are described completely by a mass parameter $M$ and a
scale radius $a$.  The Hernquist profile has density and gravitational
potential
\begin{equation}
\rho(r)=\frac{M}{2\pi}\frac{a}{r(r+a)^3},\hspace{0.5in}
\phi(r)=-\frac{GM}{r+a},
\end{equation}
while the NFW profile is given by
\begin{equation}
\rho(r)=\frac{M}{4\pi}\frac{1}{r(r+a)^2},\hspace{0.5in}
\phi(r)=-\frac{GM}{r}\ln\left(1+\frac{r}{a}\right).
\end{equation}
The Hernquist profile admits an analytic distribution function, but the NFW
profile is a better fit to dark matter halos at large radii.  We fit these
profiles to our halo by matching the observed rotation curve of the galaxy, and
we also demand that the two models have the same coefficient to the $1/r$ core
density profile.  For the Hernquist profile we find that
$M_{12}=M/(10^{12}M_\odot)=3.78$ and $a_{\rm kpc}=a/(1\;{\rm kpc})=67.1$.  The
NFW profile has $M_{12}=3.21$ and $a_{\rm kpc}=43.8$.

We now define some useful quantities.  We take ${\cal E}=\Psi(r)-v^2/2$ and
$\Psi=-\phi$.  We define dimensionless versions of the energy and potential,
$\tilde{\cal E}=q^2=a{\cal E}/(GM)$ and $\tilde{\Psi}=a\Psi/(GM)$.  We also
need a dimensionless density, which we define as $\tilde{\rho}=a^3\rho/M$.

Given that the sun moves through the halo at a velocity of 220 km sec$^{-1}$ in
the so-called ram direction, we need to know the distribution of relative
velocities between clumps and the solar system as a function of the angle from
the ram direction.  Starting from the distribution function $F$, defined so
that the mass in a phase space element is $dm=Fd^3{\bf x}\,d^3{\bf v}$, we find
that the number flux at the solar system is given by
\begin{equation}
\frac{\partial{\cal N}}{\partial v\partial\Omega}=
\frac{1}{m_{\rm clump}}Fv^2v_{\rm rel},
\end{equation}
where $v_{\rm rel}=(v^2+v^2_\odot-2vv_\odot\mu)^{1/2}$, $\theta$ is the angle
between the ram direction and the path of a given encounter, and
$\mu=\cos\theta$.  We also define $m_6=m_{\rm clump}/(10^6M_\odot)$.  Note that
a head-on collision has $\mu=-1$.  We now transform the distribution functions
to the solar system frame.  To go from the relative velocity $v_{\rm rel}$ to
the halo velocity $v$ we must choose the correct sign.  We find
$v=v_\odot\mu\pm[v_{\rm rel}^2-(1-\mu^2)v^2_\odot]^{1/2}$, and we must always
take the positive root when $v_{\rm rel}>v_\odot$ to ensure that $v>0$.
Defining $v_{100}=v_{\rm rel}/(100\,\mbox{km s$^{-1}$})$, we find that in the
frame of the solar system, the number flux is
\begin{equation}
\frac{\partial^2\cal{N}}{\partial v_{100}\partial\Omega}=
\frac{3.25\times 10^{-5}}{({\rm pc}^2\;{\rm Gy}\;{\rm sr})}
fM_{12}^{-1/2}m_6^{-1}a_{\rm kpc}^{-3/2}v_{100}^2
\left(\frac{v_{100}^2+4.84(2\mu^2-1)}{\sqrt{v_{100}^2-4.84
(1-\mu^2)}}+4.4\mu\right){\cal F}(q).
\label{distribution}
\end{equation}
The dimensionless part of the Hernquist distribution is
\begin{equation}
{\cal F}(q)=(1-q^2)^{-5/2}\left[3\sin^{-1}q+q\sqrt{1-q^2}(1-2q^2)(8q^4-8q^2-3)
\right]
\end{equation}
and $q$ is evaluated at the solar circle.  The NFW profile does not have an
analytic distribution function, but it can be evaluated numerically according
to \cite{binney},
\begin{equation}
{\cal F}(q)=
\int_0^{\tilde{\cal E}}\frac{d^2\tilde{\rho}}{d\tilde{\Psi}^2}
\frac{d\tilde{\Psi}}{(\tilde{\cal E}-\tilde{\Psi})^{1/2}}.
\end{equation}
The quantities $\tilde{\rho}$ and $\tilde{\Psi}$ are both known in terms of the
parameter $u=r/a$, $\tilde{\rho}=u^{-1}(1+u)^{-2}$ and
$\tilde{\Psi}=u^{-1}\ln(1+u)$.  Again, $q$ is evaluated at the solar circle.

We need to choose a reasonable size for the clumps.  The galaxy would process
an input spectrum of clump densities by disrupting all of the diffuse clumps,
but simple calculations indicate that an overdensity of a factor of two over
the total local density is easily safe from disruption \cite{tidal}.  We find
that the size of a clump overdense by a factor of $A$ is $r_{\rm clump}=320
(m_6/A)^{1/3}$ pc.  We use $\pi r_{\rm clump}^2$ as the cross sectional area
for purposes of determining the interaction rate.  We define the effective
interaction duration $T_{\rm int}$ as the time it takes the clump to cross the
solar system at a velocity of 400 km s$^{-1}$ times the overdensity $A$.  We
can use this quantity in Table~I to determine the integrated signal of the
encounter.  We find $T_{\rm int}=1.57 (m_6A^2)^{1/3}$ My.

We can now study the distribution of clump velocities as a function of
direction.  Multiplying equation (\ref{distribution}) by the cross sectional
area, we find the rate in terms of $A$.  We set $m_6=f=1$, $A=10$, and plot the
distribution for four values of $\mu$ in Figure~1.  In the forward direction,
the peak of the distribution is at about 450 km s$^{-1}$, and as $\mu$
increases from $-1$ to 0, the position of this peak drops to about 300 km
s$^{-1}$.  The characteristic interaction time at the peak is roughly 25
$(m_6A^2)^{1/3}f^{-1}$ My in the range $-1<\mu<0$.

The rate at which mica accumulates defects is a function of the angle $\alpha$
the incident particle makes with the cleavage plane.  Monte Carlo simulations
\cite{direction} of this process show that the rate of track accumulation
$dN/dt$ is well fit by the function
\begin{equation}
\frac{dN(\alpha)}{dt}=c_0+c_1|\cos\alpha|.
\end{equation}
Experimental measurements using neutrons to simulate WIMPs are consistent with
the etching model of the mica \cite{micadata}.  The constants $c_0$ and $c_1$
depend on the WIMP mass, the dispersion velocity of the halo, and the velocity
of the earth through the halo.  They are calculated assuming that the mica has
fixed orientation relative to the WIMP ``wind''.  We are concerned with the
signal contrast, defined as $s=(N_{\rm max}-N_{\rm min})/(N_{\rm max}+N_{\rm
min})$, where the minimum and maximum $N$ values are taken over all mica
orientations on the earth.  It is clear that the maximum possible signal
contrast $\omega$ is given by $\omega=(1+2c_0/c_1)^{-1}$, though this can only
be achieved if the mica has a fixed orientation over time.  These constants
have been computed for the standard halo \cite{direction}.  To compute the
relevant rates for the solar system interacting with a clump, we simply replace
the virial and cutoff velocities of the halo with those of the clump.  The
velocity of the sun through the halo is replaced with the relative velocity in
the interaction.  We have computed these coefficients for several interaction
velocities, in all cases taking the WIMP mass to be 100 GeV.  The values are
tabulated in Table~I along with the results for an isothermal halo.  The values
all assume a density of $0.3$ GeV cm$^{-3}$, so they must be weighted
appropriately by the density of the clump.  These results are insensitive to
the virial and cutoff velocities because the velocities of the collisions we
consider are much higher than the thermal velocities.

We need to account for the motion the mica sample makes through the galaxy.
The rotation of the earth with a period of one sidereal day and the precession
of the equinoxes with a period of 25,800 years are both very rapid compared to
any timescale we are concerned with, so we average over these motions.  The two
motions that we are more concerned with are the galactic orbit and tectonic
drift.  The galactic year is about 225 My, and the motion of the continents
occurs on timescales of 10-100 My.  All of these effects are taken into account
in computing the total track accumulation in a sample of mica as a function of
its age, continent, and orientation.  The remaining directional signal from the
isothermal halo is typically at the 1\% level \cite{direction}.

We now discuss the effect of a clump encounter at some time in the past on
the observed track density.  As our example, we consider mica that is 440 My
old, two galactic years.  Clearly we have some freedom in choosing the
parameters $m_6$, $A$, and $f$ to give an interaction rate of this order.
Conversely, we can also sample mica at different ages to roughly measure the
interaction rate.  Realistically, we can cover two orders of magnitude in
interaction rate by choosing mica of ages 10-1000 My.

We compute the track accumulation in the sample for the standard halo.  We then
compute the track accumulation due to a single $10^6 M_\odot$ clump interacting
at various times between the mica formation and the present.  Because the clump
encounter happens quickly, we may neglect the effects of galactic orbit and
tectonic drift during the encounter, but we must still average over the diurnal
rotation and the precession of the equinoxes.  When properly weighted, we can
simply add on the contribution of the nearly instantaneous clump encounter to
the isothermal result to produce the final map.  We can predict the signal
contrast and also the directions of minimum and maximum track accumulation.  We
take a conservative approach and add the clump encounter to a standard halo of
full density.  This illustrates that the clump encounters can be seen even if
they make up only a small fraction of the mass in the halo.  For $A=10$, the
change in signal can be substantial.  Some of the forward solid angle sees a
significant change in signal, but a large part sees only a small difference.
At $A=100$ the signal is much more promising.  Most of the forward solid angle
shows a pronounced change in the signal, both in that the signal contrast is
several times larger, and in that the maximum and minimum directions are very
different than the no encounter case.  When we take $A=1000$ the difference in
signal is quite large.  The signal contrast is as much as ten times larger than
the no interaction case.  The signal contrasts are listed in Table~II, and we
plot some example sky maps for mica located near the center of the North
American craton to illustrate these differences in Figure~2.  In the table and
plots we assume that the isothermal signal is at full strength, thus the signal
can only improve as $f$ increases from zero.  We have noticed that the signal
from a clump encounter 440 My ago seems to be oriented in the same direction
for all incident directions.  This is due to the fact that the North American
craton was on the equator at this time, and the signal can have no
dependence on the right ascension. For encounters occurring 220 My ago, this
degeneracy is broken since the continent has moved away from the equator.

The problem of recovering the details of a clump encounter is very much
overconstrained.  We can sample mica of different ages, different orientations,
and different continents to obtain a large amount of information about the
encounter.  The discovery of a signal can be verified with a high degree of
confidence by using other mica samples.

In summary, we stress the main point of this Letter.  Using mica as a WIMP
detector has the two advantages of direction sensitivity and integration times
as long as 1 Gy.  These two properties can be used to discover substructure in
the cold dark matter halo of the galaxy.  The hierarchical clustering paradigm
of structure formation in the universe suggests that substructure may well
persist in the halo, so this is an important experimental program.  In fact,
substructure in the halo may make unambiguous detection of halo WIMPs easier
because of the larger directional signal provided.

We would like to thank J. Silk and A. Burkert for useful discussions and
comments on the manuscript.  This research was supported in part by grants from
NASA and DOE.

\newpage
\setlength{\parindent}{0cm}
\begin{figure}[h]
\caption{Observed number flux of clumps interacting with the earth as a
function of arrival direction and velocity.  We take $m_6=f=1$ and $A=10$.
Both Hernquist and Navarro, Frenk, and White profiles are shown.  It is clear
that there is very little difference between the two.}
\label{fig:clump}
\end{figure}
\centerline{\epsfig{file=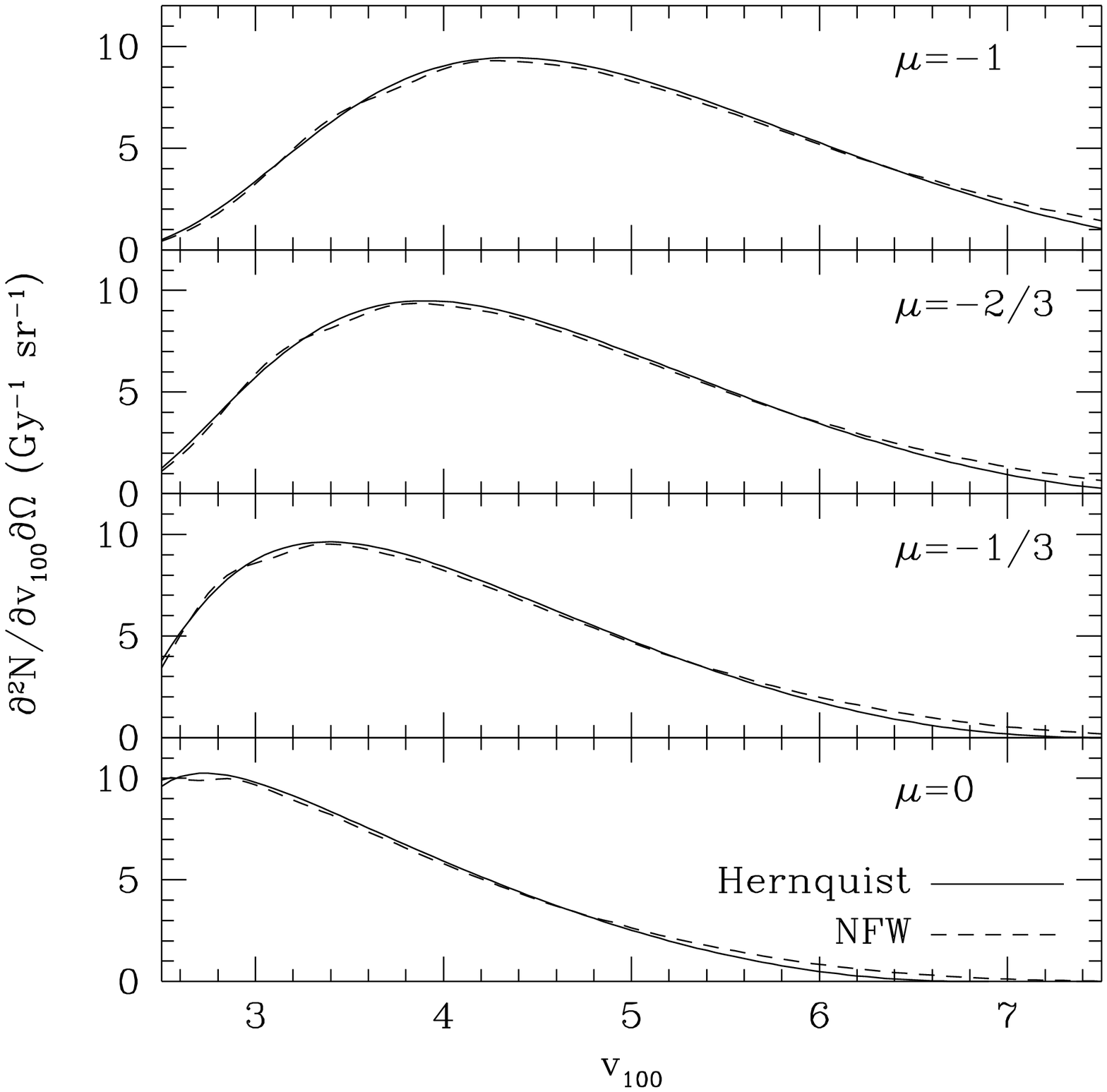}}

\begin{figure}[h]
\caption{Track accumulation as a function of present day mica orientation.  We
  illustrate the background signal (dashed) along with four clump encounters at
  various ages (solid). The background signal contrast of 0.0291 can be
  compared with the signal in each plot. {\em Top left:} $A=10$, 440 My, ram
  direction.  {\em Top right:} $A=100$, 220 My, 60 degrees north of the ram
  direction.  {\em Bottom left:} $A=100$, 275 My, 30 degrees south of the ram
  direction.  {\em Bottom right:} $A=1000$, 275 My, 60 degrees west of the ram
  direction.}
\label{fig:map}
\end{figure}
\epsfig{file=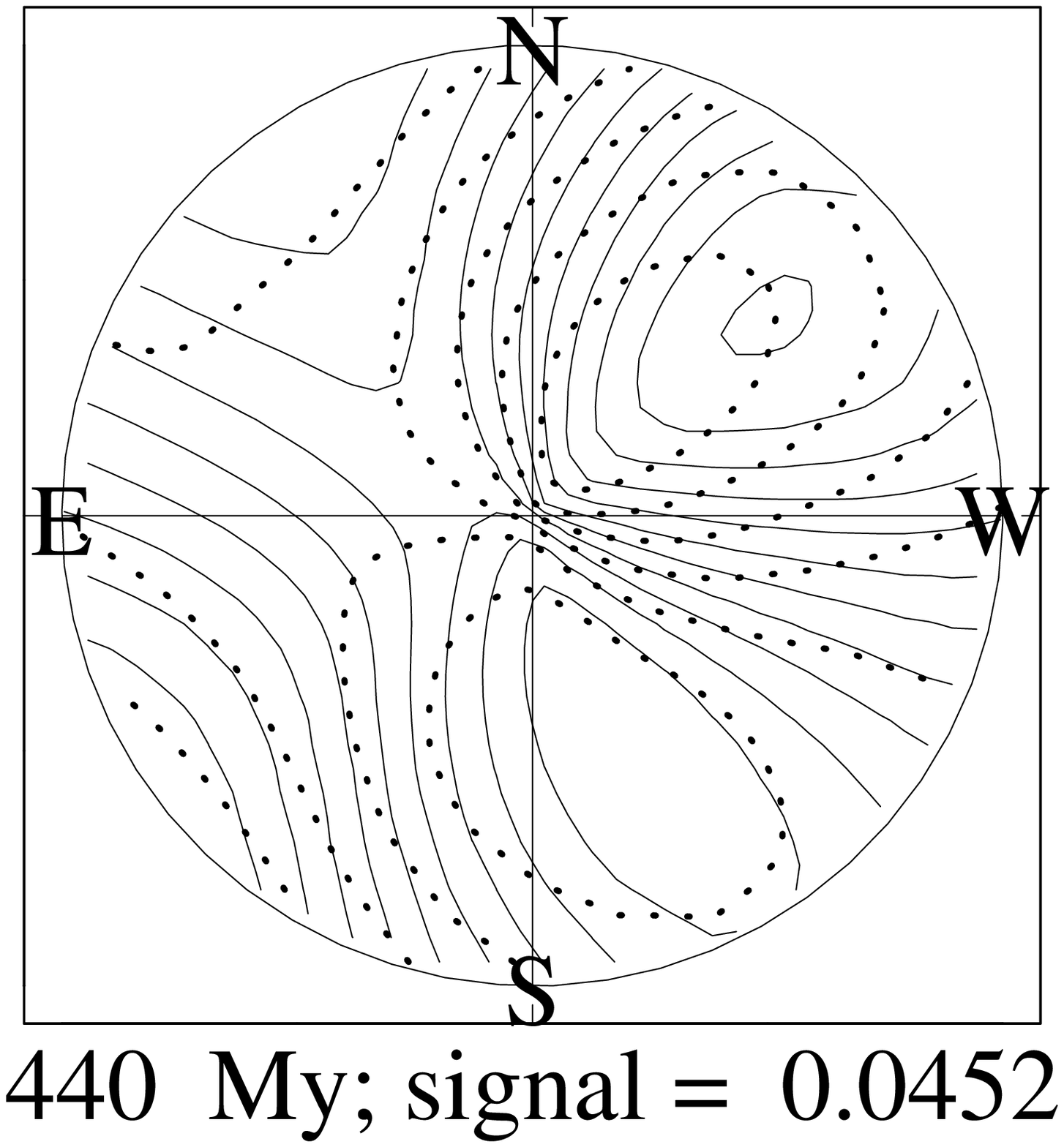,width=3in}
\epsfig{file=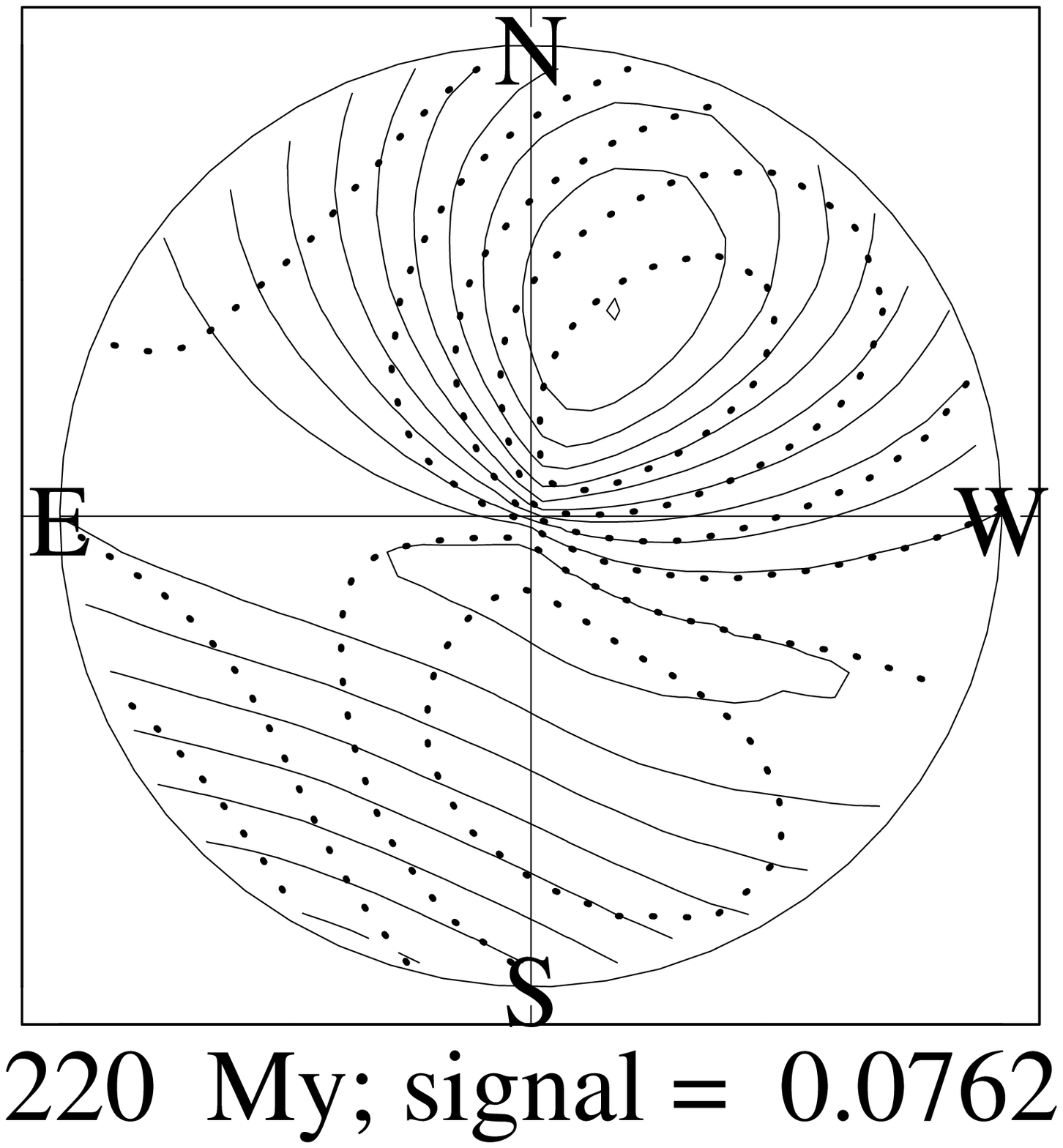,width=3in}\vspace{-1in}
\\ \epsfig{file=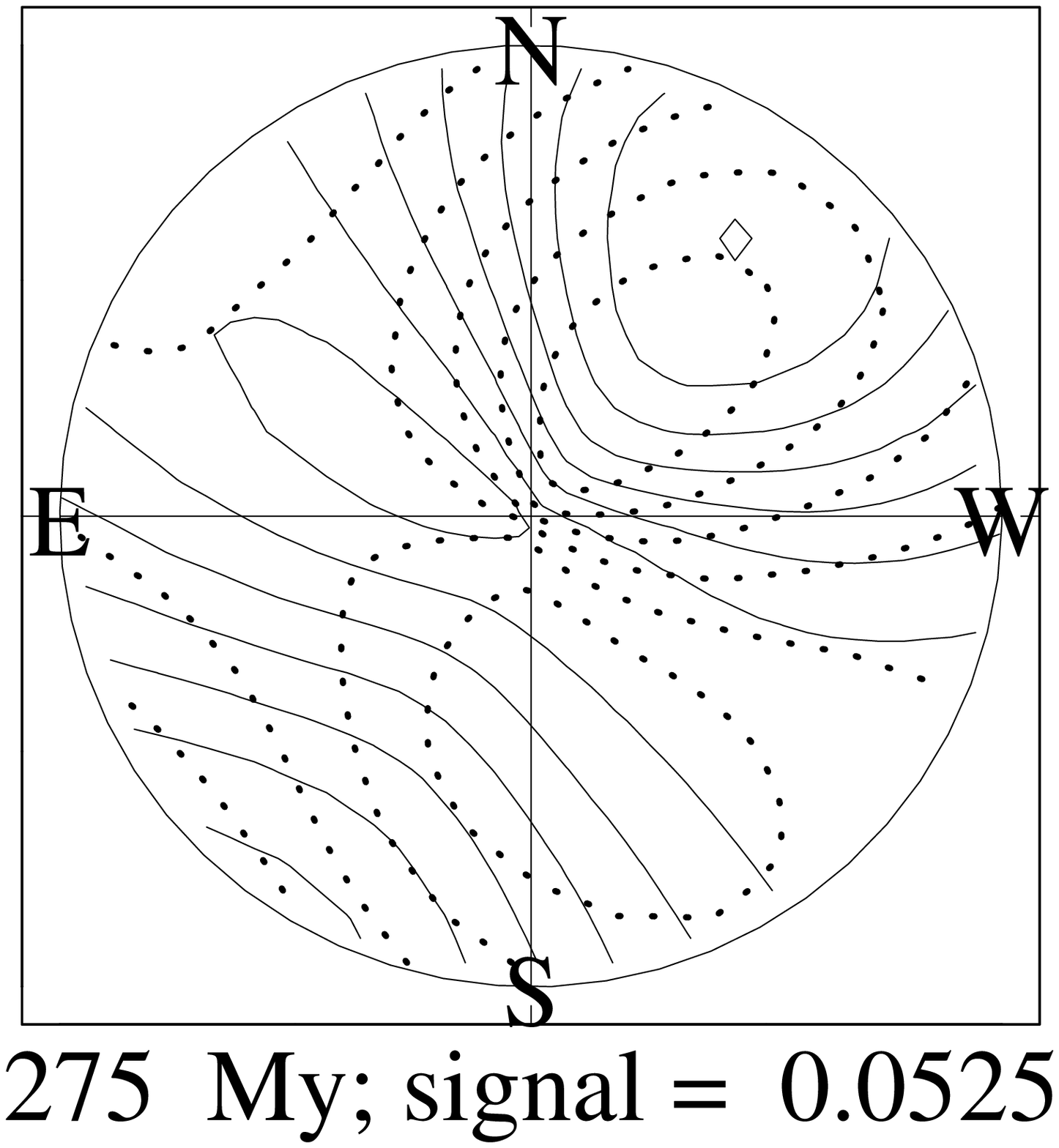,width=3in}
\epsfig{file=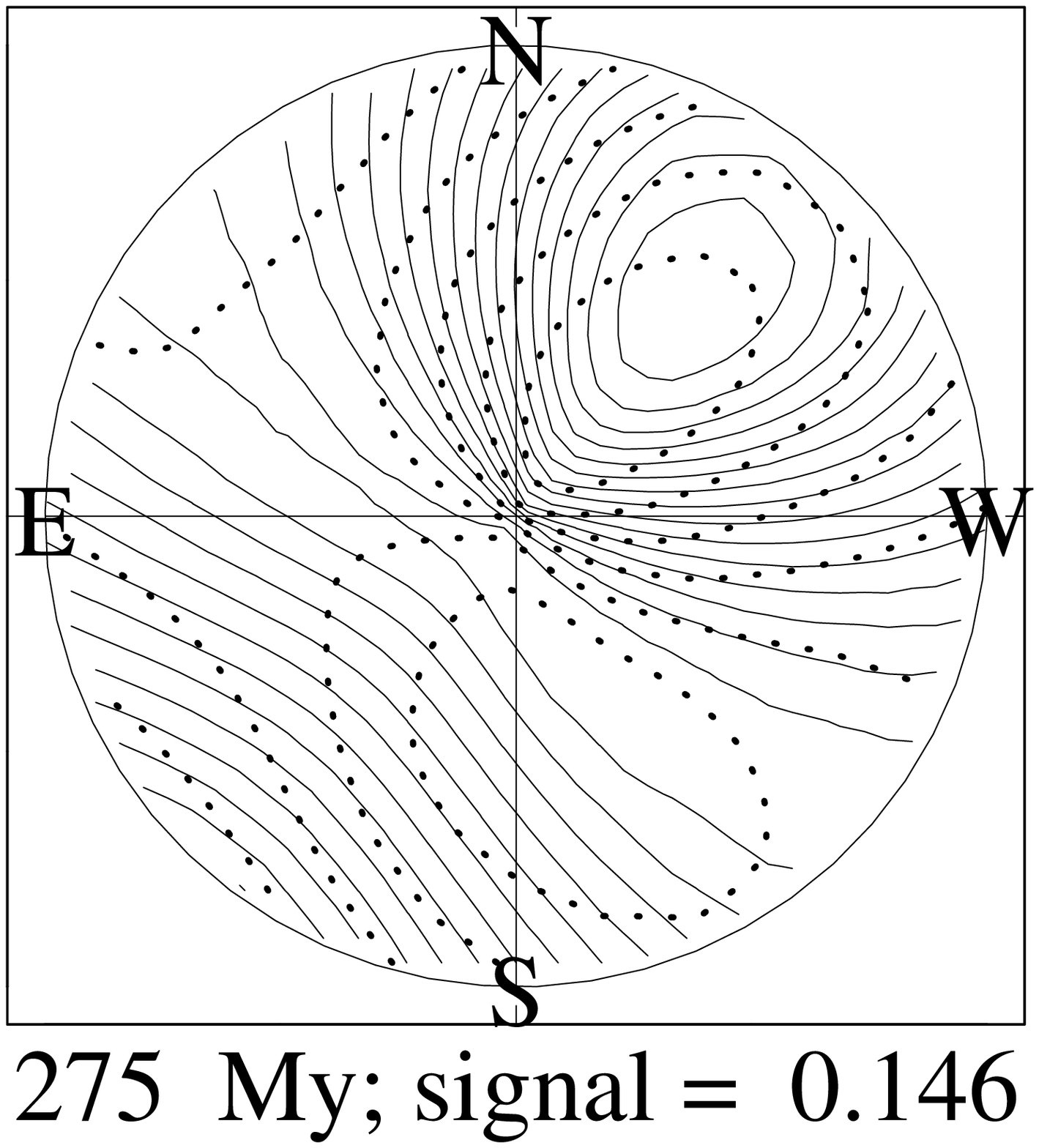,width=3in}

\newpage
\begin{table}
\mediumtext
\caption{Track accumulation rate coefficients $c_0$ and $c_1$ and signal
contrast $\omega$ for several interaction velocities.  The values for the
standard isothermal halo are also given.  The WIMP mass is taken to be 100
GeV.}
\begin{tabular}{cr@{${}\pm{}$}rr@{${}\pm{}$}rd@{${}\pm{}$}d}
Velocity & \multicolumn{2}{c}{$c_0$} & \multicolumn{2}{c}{$c_1$} &
\multicolumn{2}{c}{$\omega$} \\
(km s$^{-1}$) & \multicolumn{2}{c}{(cm$^{-2}$ Gy$^{-1}$)} &
\multicolumn{2}{c}{(cm$^{-2}$ Gy$^{-1}$)} & \multicolumn{2}{c}{}\\
\hline
300 & -16 700 & 43 000 & 370 000 & 61 000 & 1.10 & 0.28\tablenote{Note that
$|\omega|>1$ is unphysical, but in this case the $1\sigma$ error bars cover
unity so we set $\omega=1$} \\
400 & 550 000 & 200 000 & 1 800 000 & 290 000 & 0.62 & 0.094 \\
500 & 4 800 000 & 360 000 & 1 500 000 & 500 000 & 0.14 & 0.04 \\
600 & 12 700 000 & 500 000 & -700 000 & 740 000 & -0.03 & 0.03 \\
700 & 27 400 000 & 960 000 & -7 300 000 & 1 400 000 & -0.15 & 0.03 \\
halo & 2 860 000 & 80 000 & 700 000 & 100 000 & 0.11 & 0.001
\end{tabular}
\end{table}

\begin{table}
\mediumtext
\caption{Signal contrast for encounters with a $10^6 M_\odot$ clump in mica
  440 My old.  The clump overdensity $A$ and encounter time in My are listed
  with the arrival direction relative to the ram direction in galactic
  coordinates.  Clumps coming from the galactic center and north galactic pole
  are also listed.}
\begin{tabular}{llllllllll}
Direction & \multicolumn{3}{c}{$A=10$} & \multicolumn{3}{c}{$A=100$} &
\multicolumn{3}{c}{$A=1000$} \\
& \multicolumn{1}{c}{220} & \multicolumn{1}{c}{275} & \multicolumn{1}{c}{440} &
\multicolumn{1}{c}{220} & \multicolumn{1}{c}{275} & \multicolumn{1}{c}{440} &
\multicolumn{1}{c}{220} & \multicolumn{1}{c}{275} & \multicolumn{1}{c}{440} \\
\hline
no clump & 0.0291 & 0.0291 & 0.0291 & 0.0291 & 0.0291 & 0.0291 & 0.0291 &
0.0291 & 0.0291 \\
ram & 0.0446 & 0.020 & 0.0452 & 0.102 & 0.0573 & 0.106 & 0.229 & 0.189 &
0.233 \\
pole & 0.0286 & 0.0284 & 0.0285 & 0.0268 & 0.0260 & 0.0265 & 0.0217 & 0.0195 &
0.0240 \\
center & 0.0274 & 0.0316 & 0.0279 & 0.0236 & 0.0412 & 0.0255 & 0.0424 & 0.0833
& 0.0497 \\
30$^\circ$E & 0.0394 & 0.0211 & 0.0388 & 0.0792 & 0.0345 & 0.0809 & 0.172 &
0.125 & 0.175 \\
30$^\circ$S & 0.0259 & 0.0201 & 0.0255 & 0.020 & 0.0525 & 0.0248 & 0.0412 &
0.177 & 0.0456 \\
30$^\circ$W & 0.0336 & 0.0252 & 0.0319 & 0.0533 & 0.0182 & 0.0522 & 0.102 &
0.0303 & 0.103 \\
30$^\circ$N & 0.0550 & 0.0209 & 0.0569 & 0.143 & 0.0368 & 0.151 & 0.332 & 0.131
& 0.342 \\
60$^\circ$E & 0.0272 & 0.0330 & 0.0268 & 0.0226 & 0.0473 & 0.0223 & 0.0261 &
0.0891 & 0.0304 \\
60$^\circ$S & 0.0237 & 0.0238 & 0.0255 & 0.0419 & 0.0204 & 0.0489 & 0.148 &
0.0807 & 0.152 \\
60$^\circ$W & 0.0250 & 0.0372 & 0.0256 & 0.0269 & 0.0662 & 0.0336 & 0.0969 &
0.146 & 0.102 \\
60$^\circ$N & 0.0380 & 0.0257 & 0.0381 & 0.0762 & 0.0190 & 0.0785 & 0.186 &
0.0366 & 0.191
\end{tabular}
\end{table}


\begin{thebibliography}{99}
\bibitem{dark}
V. Trimble, Ann. Rev. Astron. Astrophys. {\bf 25}, 425 (1987)

\bibitem{susydm}
G. Jungman, M. Kamionkowski, and K. Griest, Phys. Rep. {\bf 267}, 195 (1996)
and references therein.

\bibitem{mica}
D. P. Snowden-Ifft, E. S. Freeman, and P. B. Price, Phys. Rev. Lett. {\bf 74},
4133 (1995).

\bibitem{etch}
D. P. Snowden-Ifft and M. K. Y. Chan, Nucl. Instrum. Methods B {\bf 101}, 247
(1995).

\bibitem{direction}
D. P. Snowden-Ifft and A. J. Westphal, Phys. Rev. Lett. {\bf 78}, 1628 (1997).

\bibitem{hierarchy}
W. H. Press and P. Schechter, Astrophys. J. {\bf 187}, 425 (1974); C. Lacey and
S. Cole, Mon. Not. R. Astron. Soc. {\bf 262}, 627 (1993); {\bf 271}, 676
(1994).

\bibitem{streamers}
K. V. Johnston, astro-ph/9710007 (1997).

\bibitem{andi}
A. Burkert and J. Silk, private communication

\bibitem{hernquist}
L. Hernquist, Astrophys. J. {\bf 356}, 359 (1990).

\bibitem{navarro}
J. F. Navarro, C. S. Frenk, and S. D. M. White, Mon. Not. R. Astron. Soc. {\bf
275}, 720 (1995); Astrophys. J. {\bf 462}, 563 (1996); (to be published)
astro-ph/9611107.

\bibitem{micadata}
D. P. Snowden-Ifft and E. S. Rykoff, unpublished

\bibitem{binney}
J. Binney and S. Tremaine, {\em Galactic Dynamics} (Princeton University Press,
Princeton, 1987).

\bibitem{tidal}
E. R. Capriotti and S. L. Hawley, Astrophys. J. {\bf 464}, 765 (1996).

\end{thebibliography}
\end{document}